\documentclass[aps,pra,reprint,superscriptaddress]{revtex4-2}
\usepackage{amsmath}
\usepackage{amssymb}
\usepackage{mathtools}
\usepackage{xcolor}
\usepackage{graphicx}
\usepackage[normalem]{ulem}

\newcommand{\ket}[1]{\left\vert#1\right\rangle}

\begin{document}

\title{Switching-free time-domain optical quantum computation with quantum teleportation}


\author{Warit Asavanant}
\email{warit@alice.t.u-tokyo.ac.jp}
\affiliation{Department of Applied Physics, School of Engineering, The University of Tokyo, 7-3-1 Hongo, Bunkyo-ku, Tokyo 113-8656, Japan}
\author{Kosuke Fukui}
\email{fukuik.opt@gmail.com}
\affiliation{Department of Applied Physics, School of Engineering, The University of Tokyo, 7-3-1 Hongo, Bunkyo-ku, Tokyo 113-8656, Japan}
\author{Atsushi Sakaguchi}
\email{atsushi.sakaguchi@riken.jp}
\affiliation{Optical Quantum Computing Research Team, RIKEN Center for Quantum Computing,\\ 2-1 Hirosawa, Wako, Saitama 351-0198, Japan}
\author{Akira Furusawa}
\email{akiraf@ap.t.u-tokyo.ac.jp}
\affiliation{Department of Applied Physics, School of Engineering, The University of Tokyo, 7-3-1 Hongo, Bunkyo-ku, Tokyo 113-8656, Japan}
\affiliation{Optical Quantum Computing Research Team, RIKEN Center for Quantum Computing,\\ 2-1 Hirosawa, Wako, Saitama 351-0198, Japan}


\date{\today}

\begin{abstract}
Optical switches and rerouting network are main obstacles to realization of optical quantum computer. Both components have been long considered as essential components to the measurement-based time-domain optical quantum computation, an approach which has seen promising developments in the recent years regarding scalability. Realizing optical switches and rerouting network with sufficient performance is, however, experimentally challenging as they must have extremely low loss, small switching time, high repetition rate, and minimum optical nonlinearity. In this work, we present an optical quantum computation platform that does not require such optical switches. Our method is based on continuous-variable measurement-based quantum computation, where instead of the typical cluster states, we modify the structure of the quantum entanglements, so that quantum teleportation protocol can be employed instead of the optical switching and rerouting. We also show that when combined with Gottesman-Kitaev-Preskill encoding, our architecture can perform particularly well compare to the architecture with optical switches when the optical losses of the switches are not low.
\end{abstract}


\maketitle



\section{Introduction}
Optical systems are promising candidates for quantum computation as they can be implemented at room temperature and atmospheric pressure, and have high compatibility with optical communications. Optical quantum computation can be largely categorized into two types: discrete variable (DV) and continuous variable (CV). Various researches have been done for both types and many different architecture have been theoretically developed and experimentally demonstrated \cite{doi:10.1063/1.5100160,doi:10.1063/1.5115814}. From these various platforms, time-domain CV measurement-based quantum computation \cite{PhysRevA.83.062314} has shown a great potential regarding scalability: large-scale universal computational resource---the cluster state \cite{PhysRevLett.86.5188,PhysRevLett.97.110501}---has been generated \cite{Yokoyama2013,doi:10.1063/1.4962732,Asavanant373,Larsen369} and basic operations on these scalable platform have also been demonstrated \cite{PhysRevApplied.16.034005,Larsen2021}.

Regardless of the details of each optical platform, optical switch is considered a key component. In the DV optical quantum computer, making a semi-deterministic single-photon source via multiplexing of probabilistic single-photon sources requires optical switches \cite{doi:10.1063/5.0003320}. Also, tunable universal linear optics, which can be realized with optical switching and phase shifters, is essential when programming the DV optical quantum computer \cite{doi:10.1126/science.aab3642}. On the other hand, for the CV platforms---in particular, the measurement-based platform---optical switches are used in tasks such as the injection and dejection of the quantum states, injection of the magic states, error-correction circuits, and multiplexing of non-Gaussian ancillary states \cite{Larsen2019,PhysRevA.97.032302,PhysRevLett.119.120504,PRXQuantum.2.030325}. Despite the current development of the optical switching for the optical quantum computer \cite{Larsen2019,doi:10.1126/sciadv.aaw4530}, simultaneously realizing extremely low optical loss, high-speed switching, and high repetition rate is highly challenging. For example, in a optical switch using pockel cell architecture which usually have low optical loss, voltage required in the switching is typically on the order of a few thousands volt, which limits the repetition rate. The repetition rate is especially important in the time-domain measurement-based architecture, where the inputs are encoded in a series of time-binned wave packets. In addition, as optical switching usually uses nonlinear optical medium, it is important to keep this nonlinearity from affecting the fragile quantum states. This can be considered challenging for some types of optical switches where high optical nonlinearity is required \cite{Yoshiki:14}.

In this work, we present a switching-free CV measurement-based optical quantum computation platform in time domain. The main idea of this platform is that we appropriately modify the entanglement structure of the cluster states so that it can also function as optical switches and rerouting which are required when using the cluster states for quantum computation. This additional entanglement structure can be used for injection of the input state, multiplexing/rerouting of the ancillary state, and coupling of the cluster states to the ancillary states in the error syndrome measurement. We can select and switch between each function by switching the measurement bases of the homodyne measurement, allowing these functions to be implemented in via quantum teleportation protocol. This removes the necessity of the inline active optical components, and the required offline active optical components are squeezed light sources and phase modulator for the local oscillators, which are both established technologies \cite{PhysRevApplied.16.034005,Larsen2021}. We also perform a numerical calculation showing that depending on the quality of the optical switches, our system can realize a lower error probability when combined with Gottesman-Kitaev-Preskill (GKP) encoding \cite{PhysRevA.64.012310}. Although a similar concept can be seen in the fusion gate of the DV optical qubits \cite{PhysRevLett.95.010501}, unlike the DV systems, the CV entanglements are generated deterministically and continuously, making our approach experimentally viable. Therefore, this architecture removes the necessity of the optical switching from fault-tolerant universal optical quantum computation platform, making it unique compared to other time-domain CV architecture \cite{Asavanant373,Larsen369,PhysRevLett.119.120504,PRXQuantum.2.030325}.  

This paper is structured as follows. Section \ref{sec:preliminaries} explains preliminaries and notations used in this paper. The proposed setup and its analysis are shown in Sec.\ \ref{sec:mainresult}. Section \ref{sect:incorportate} describes how to take into account of photon loss in optical switch when for both GKP states and cluster states. Section \ref{sec;analysis} shows the analysis of the error probability of both conventional method and our method. Section \ref{sec:numerical} discusses the experimental feasibility and shows the numerical analysis. Section \ref{sec:conclusion} concludes the paper.

\section{Preliminaries}\label{sec:preliminaries}
First we explain the basic notations and review the concepts of CV measurement-based quantum computation.

\subsection{Notations}
In the CV quantum computation, the physical quantities of our interest are quadratures. Quadratures are denoted by operators $\hat{x}$ and $\hat{p}$ which satisfy $[\hat{x},\hat{p}]=i$ (corresponds to $\hbar=1$). The quadrature operators are related to annihilation operator $\hat{a}$ and creation operator $\hat{a}^\dagger$ via:
\begin{align}
\hat{x}&=\frac{1}{\sqrt{2}}(\hat{a}+\hat{a}^\dagger),\\
\hat{p}&=-\frac{i}{\sqrt{2}}(\hat{a}-\hat{a}^\dagger).
\end{align}

Next, we define some of the basic operations in this paper. First, a squeezing operator $\hat{S}(r)$ with a squeezing parameter $r$ transforms the quadrature operators in Heisenberg picture as
\begin{align}
\hat{S}^\dagger(r)\begin{pmatrix}
\hat{x}\\
\hat{p}
\end{pmatrix}\hat{S}(r)=\begin{pmatrix}
\hat{x}e^{r}\\
\hat{p}e^{-r}
\end{pmatrix}.
\end{align}
Another important operation is a phase rotation $\hat{R}(\theta)$ which transforms quadrature operators as
\begin{align}
\hat{R}^\dagger(\theta)\begin{pmatrix}
\hat{x}\\
\hat{p}
\end{pmatrix}
\hat{R}(\theta)=\begin{pmatrix}
\cos\theta&\sin\theta\\
-\sin\theta&\cos\theta
\end{pmatrix}\begin{pmatrix}
\hat{x}\\
\hat{p}
\end{pmatrix}.
\end{align}

Regarding two-mode operation, we first consider beamsplitter interaction $\hat{B}_{12}(\sqrt{R})$ which we define as
\begin{multline}
\hat{B}_{12}^\dagger(\sqrt{R})\begin{pmatrix}
\hat{x}_1\\
\hat{x}_2\\
\hat{p}_1\\
\hat{p}_2
\end{pmatrix}\hat{B}_{12}(\sqrt{R})=\\
\begin{pmatrix}
\sqrt{R}&\sqrt{T}&0&0\\
-\sqrt{T}&\sqrt{R}&0&0\\
0&0&\sqrt{R}&\sqrt{T}\\
0&0&-\sqrt{T}&\sqrt{R}
\end{pmatrix}\begin{pmatrix}
\hat{x}_1\\
\hat{x}_2\\
\hat{p}_1\\
\hat{p}_2
\end{pmatrix}.
\end{multline}
Another two-mode operation which will be used in this paper is a CV version of the controlled-Z gate, denoted by $\hat{C}_Z$. We will consider a control-Z gate with gain of 1, which transform the quadrature operators as
\begin{align}
\hat{C}_Z^\dagger\begin{pmatrix}
\hat{x}_1\\
\hat{x}_2\\
\hat{p}_1\\
\hat{p}_2
\end{pmatrix}
\hat{C}_Z=
\begin{pmatrix}
1&0&0&0\\
0&1&0&0\\
0&-1&1&0\\
-1&0&0&1
\end{pmatrix}\begin{pmatrix}
\hat{x}_1\\
\hat{x}_2\\
\hat{p}_1\\
\hat{p}_2
\end{pmatrix}.
\end{align}

\subsection{Quantum entanglement and quantum computation}

It is known that quantum entanglements with appropriate structures, i.e., the cluster states, allow universal quantum computation when combined with local (single mode or single qubit) measurements and feedforward operations that depend on the measurement results \cite{PhysRevLett.86.5188,PhysRevLett.97.110501}. Quantum computation using cluster states can be equivalently considered as sequential quantum teleportation where the measurement bases are chosen depending on the desired operations \cite{Yokoyama2013}.

Quantum entanglement (and pure Gaussian states in general) in CV system is defined by a set of operators called nullifiers. For a $N$-mode pure Gaussian state $\ket{G}$, an operator $\hat{\delta}$ is a nullifier of $\ket{G}$ if and only if
\begin{align}
\hat{\delta}\ket{G}=0
\end{align}
and the state $\ket{G}$ can be uniquely defined by $N$ independent nullifiers.

In the generations of many ideal CV quantum entanglement, including the cluster states, infinite squeezing is required. In the actual physical situations, however, we can only achieve finite squeezing. The imperfections due to the finite squeezing appear as the non-zero variances of the nullifiers. When CV quantum entanglements are used, especially in measurement-based quantum computation, the squeezing of the nullifiers corresponds to the additional noises to the output, even when the quantum entanglement is not a pure Gaussian state \cite{PhysRevA.100.010301}. The reason behind this is that the effects of the anti-squeezing components are erased by the feedforward operations. Therefore, when considering the CV quantum entanglements and their usage in measurement-based quantum computation, the variances of the nullifiers are the primary figures of merit.

\subsection{Time-domain multiplexing method and macronodes}
\begin{figure}[htb]
\includegraphics[width=\columnwidth]{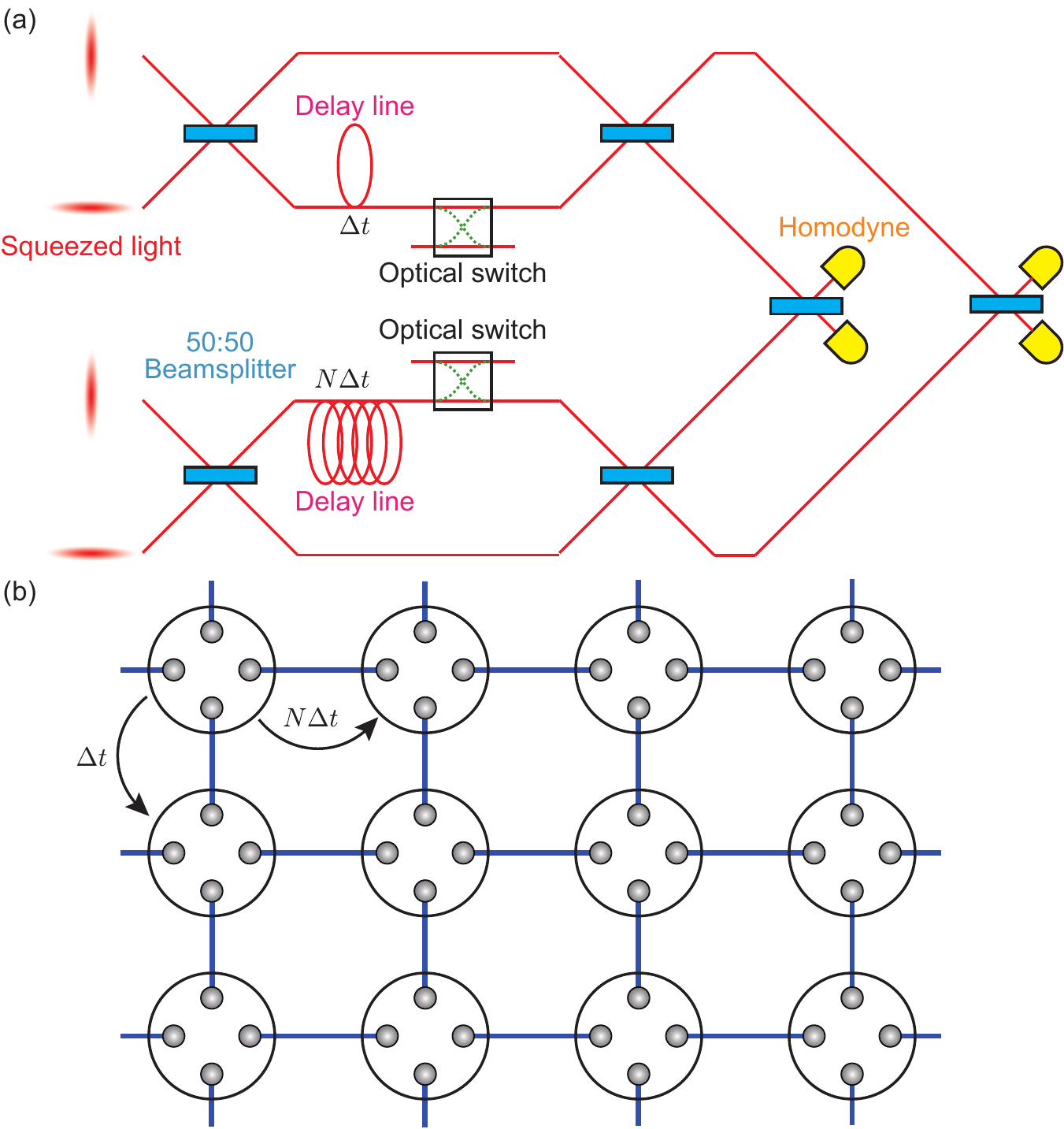}%
\caption{Schematic diagram of time-domain-multiplexed quad-rail lattice cluster state setup. (a) The setup which consists of four squeezed light sources, five beamsplitters, four homodyne detectors, and optical switches. (b) Macronode representation of quad-rail lattice cluster state where the two-mode entanglement shown are two-mode squeezed state (EPR state in the infinite squeezing limit). Note that when considering quantum error correction, a switching for implementation of error correction is also required but it is omitted here \protect\cite{PhysRevA.102.042608}.\label{fig:QRL_setup}}
\end{figure}
Figure \ref{fig:QRL_setup} shows an experimental setup of the time-domain measurement-based quantum computation using a type of two-dimensional cluster states called quad-rail lattice \cite{PhysRevA.83.062314}. In time-domain quantum computation, the modes of the cluster states are defined as a localized temporal wave packets in a time bin. By using two optical delay lines and linear optics, these wave packets are interfered resulting in the two-dimensional cluster state. 

Although the quad-rail lattice cluster state can be reduced to conventional square-lattice cluster state via measurements and feedforward operations, a more efficient approach can be achieved by introducing a concept of macronode; instead of the cluster state with complex structures, the operations on the quad-rail cluster state are equivalent to quantum teleportation where the two-mode squeezed states are measured using a network of linear optics and homodyne measurement \cite{PhysRevA.93.062326}. The same concept is also applied to two-dimensional cluster states with different structures to achieve efficient quantum computation \cite{PhysRevA.102.042608}. In the specific case of the quad-rail lattice cluster state, by choosing appropriate measurement bases, we can select between implementation of one-mode unitary operation on each input mode or interaction between two modes.

Although quantum computation using DV system does not require external input states as the nodes of the cluster states can be initialized as the inputs, for the CV state, the external input states is preferable as it usually requires too many steps to make a complex state. Injection of external quantum states into the CV cluster state requires optical switch as shown in Fig.\ \ref{fig:QRL_setup}. In addition to the state injection, we also need the switching for the ancillary state used in error syndrome measurement and non-Gaussian measurement \cite{PhysRevA.97.032302,PhysRevA.102.042608}. These optical switches must have extremely low optical losses, high switching speed, and high repetition rate to be compatible with the time-domain multiplexing method.

\section{Main results}\label{sec:mainresult}

\subsection{Proposed setup}
\begin{figure}[htb]
\includegraphics[width=\columnwidth]{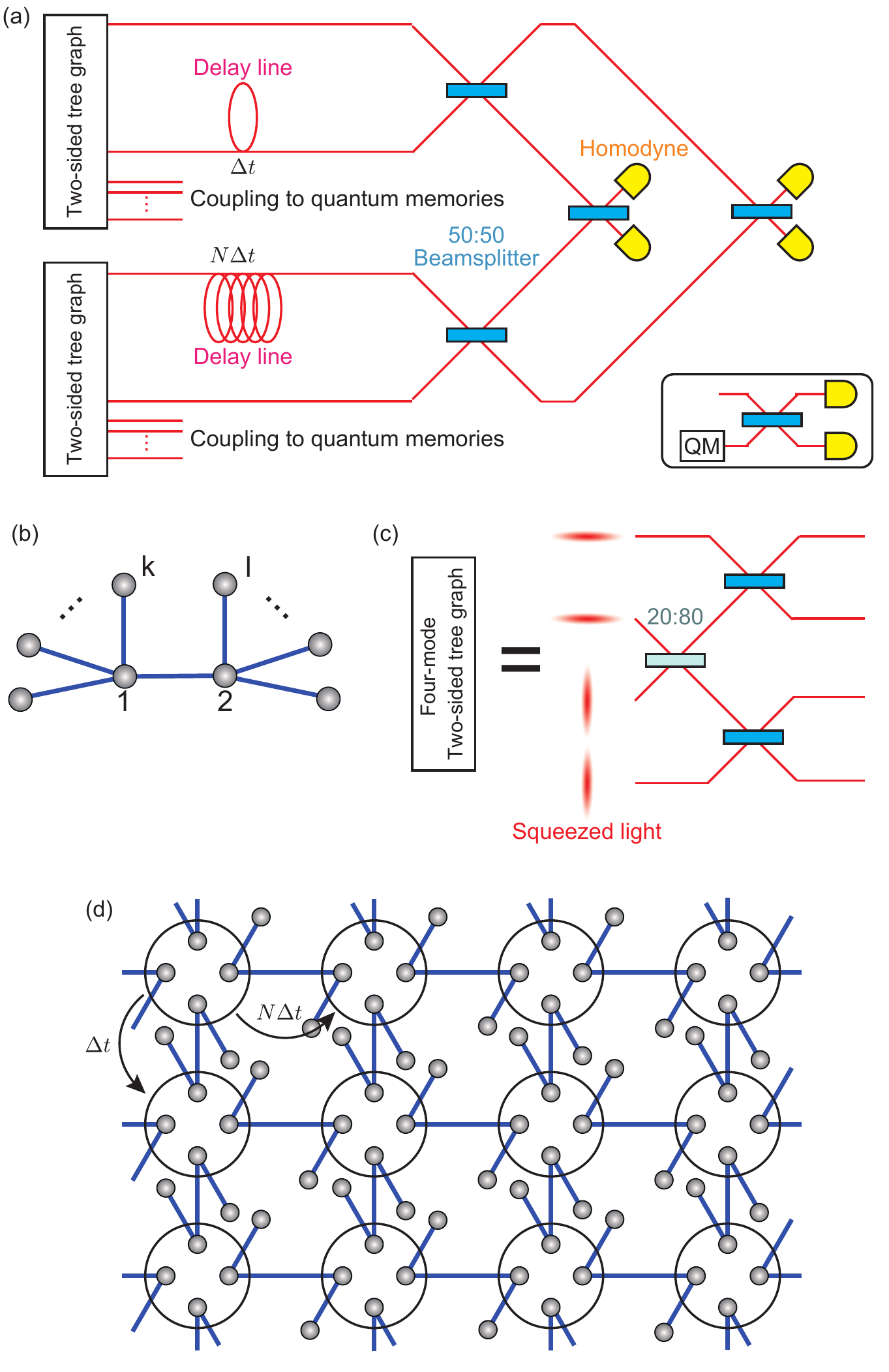}%
\caption{Proposed switching-free optical quantum computer setup. (a) Experimental setup. The inset shows the coupling of the mode of the entanglement to the quantum memory (QM).  (b) Two-sided tree graph state. (c) An example of the generation setup for the four-mode two-sided tree graph which is equivalent to four-mode linear cluster state. (d) Macronode representation of the setup. For simplicity, we show the case with four modes. \label{fig:setup}}
\end{figure}
To circumvent the needs of the optical switching which is a current bottle neck to the development of the optical quantum computer, we propose a setup shown in Fig.\ \ref{fig:setup}. In this setup, we replace the two-mode squeezed states in Fig.\ \ref{fig:QRL_setup} with a CV quantum entanglement that has a tree-graph structure out branching out at both ends Fig.\ \ref{fig:QRL_setup}(b)]---which we will call \emph{two-sided tree graph}. The nullifiers of the two-sided tree graph is given by
\begin{align}
\hat{\delta}_1&=\hat{p}_1-\hat{x}_2-\sum_{k\in\mathcal{K}}\hat{x}_{k},\\
\hat{\delta}_2&=\hat{p}_2-\hat{x}_1-\sum_{l\in\mathcal{L}}\hat{x}_{l},\\
\hat{\delta}_{k}&=\hat{p}_k-\hat{x}_1,\\
\hat{\delta}_{l}&=\hat{p}_{l}-\hat{x}_2,
\end{align}
where $\mathcal{K}$ ($\mathcal{L}$) is a set of nodes that is connected to mode 1 (2) and does not include mode 2 (1). Note that it is possible to transform this state into a state whose nullifiers composed of only quadrature operator $\hat{x}$ or $\hat{p}$. Such type of state is the called $\mathcal{H}$-graph which can be generated via appropriate generalized parametric down conversion.

The branches of the two-sided tree graph are then coupled to a quantum memory via a 50:50 beamsplitter, where the mode 1 and 2 are inside the cluster circuit and is used for operations. Figure \ref{fig:setup}c shows an example of a possible generation setup of the two-sided tree graph in the four-mode case. This is the case with smallest number of modes and the resource state become a four-mode linear cluster states which have already been experimentally demonstrated \cite{PhysRevA.78.012301}. For a case with more modes, it is theoretically shown that arbitrary Gaussian states can be generated with offline squeezing and linear optics \cite{PhysRevA.76.032321}, meaning that we can definitely find a physical circuit that realizes this initial resource state.
 
\subsection{Input state injection and rerouting}
In optical quantum computation, quantum states besides Gaussian states are usually generated using heralding method. Due to the random generation timing in the heralding method, the state generation is usually combined with quantum memories (For example, \cite{PhysRevLett.123.113603}). Then, optical switches are used to inject the generated states from the quantum memory into the cluster states. To increase the generation rate, it is expected that a network of quantum memories and rerouting system using optical switching would be required. Realizing this rerouting system with low optical losses is, however, technically difficult. Also, if the rerouting is slow, the required storage time of the optical memory will be longer which would degrade the state even more. 

\begin{figure}
\includegraphics[width=\columnwidth]{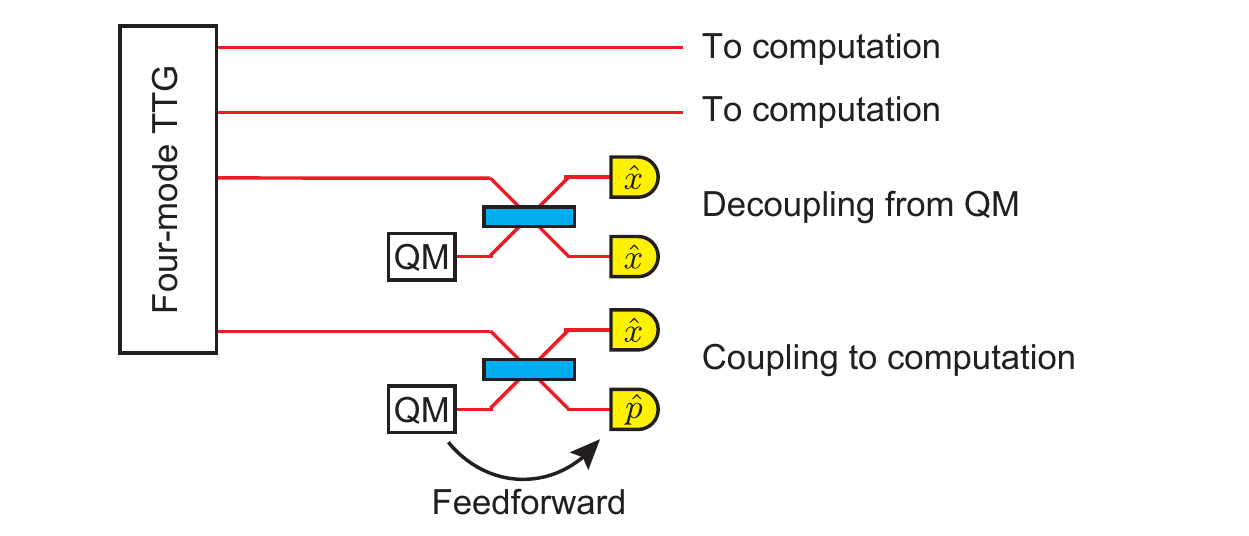}%
\caption{State injection and rerouting with two-sided tree graph state. Here we show the case with four-mode two-sided tree graph (TTG), but this can be made arbitrary. When the state is generated and successfully stored in the quantum memory (QM), this information is used to feedforward to the measurement bases of the homodyne, and the quantum state is teleported as the input of our computation. The other quantum memories are decoupled to the computation by homodyne measurement of the same bases. Here we omitted the displacement operations that depend on the measurement results.\label{fig:rerout}}
\end{figure}

Here, we discuss how our proposed setup removes the need of the optical switches and rerouting network. The main idea is illustrated in Fig.\ \ref{fig:rerout} for a case with two quantum memories. First, two of the modes of the entanglement (the other two modes are used in the computation) are distributed and interfere on a 50:50 beamsplitter which is with the beam from the quantum memory, and a homodyne detector is put each output port of the beamsplitter. When we do not need to inject the quantum state, the measurement bases of all the homodyne detectors are set to $\hat{x}$. When the quantum state we wish to inject is successfully generated and stored inside one of the quantum memory, we use the quantum memory to adjust the timing, shift the measurement basis of one of the homodyne detectors to $\hat{p}$, and teleport the state into the setup for further computation. The degree of freedom at the homodyne detectors for state injection can also be used to implement additional Gaussian operations if required \cite{PhysRevA.81.032315}. Note, however, that the quantum memory is required when we consider continuous-wave optical generation of the non-Gaussian state, where the state generations occur at a random timing. For a system where the generation timing is well-defined such as pulsed-laser system, we only require a fixed delay line to adjust the timing of feedforward operation. Comparing to the rerouting network, switching of the homodyne measurement basis is much easier as this is done by phase shifting of the classical local oscillator, meaning that we do not have to care about the optical losses in this case. Do note, however, that there will be a trade-off relation between the number of branches and noises property as actual CV quantum entanglement can only be generated with finite squeezing. We will discuss this in further details in Sec.\ \ref{sec:numerical}. Similar to the state injection, we can also deject the state from the cluster state by using the branch of the additional entanglement as the output mode.

\subsection{Universality}
We also show the universality of our setup. In particular, universal operations on the GKP qubit is an important aspect as the GKP qubit is a promising encoding for universal and fault-tolerant quantum computation with continuous variables \cite{PhysRevLett.112.120504,PhysRevLett.123.200502,PhysRevLett.125.040501,PhysRevLett.119.180507,PhysRevX.8.021054,2019arXiv190609767F,PhysRevA.102.062411,PhysRevA.101.012316}. Since the two-sided tree graph state can be reduced to a two-mode cluster state, the implementation of the Gaussian operations follows the protocol in Ref.\ \cite{PhysRevA.93.062326} and can be done with homodyne measurements and feedforward operations.

On the other hand, there are several ways non-Gaussian operations can be implemented. The most basic way would be by replacing one of the homodyne detectors with nonlinear measurement such as cubic phase measurement \cite{PhysRevA.97.032302}. This, however, requires optical switching to switch between the homodyne measurement and cubic phase measurement. As the cubic phase measurement utilize Gaussian operations and non-Gaussian ancillary state, we could instead use the entanglement structure of the two-sided tree graph and coupling the non-Gaussian ancillary states to the input states and then follow the measurement-based methodology to implement non-Gaussian gates \cite{PhysRevA.97.022329}.

If we consider GKP encoding, the Clifford operations can be implemented using Gaussian operations. On the other hand, non-Clifford operations required non-Gaussian operations \cite{PhysRevA.64.012310}. Recently there has been a proposal where GKP non-Clifford operation is efficiently implemented using nonlinear feedforward system and ancillary state \cite{PhysRevResearch.3.043026}. We can use the entanglement structure of the two-sided tree graph to couple the input with the ancillary state here and implement the adaptive homodyne measurement which results in the non-Clifford gate for GKP qubits after the feedforward operations. In this sense, the branches that are used for injection of the quantum state can also be used in the implementation of the non-Clifford operations.

\subsection{Error syndrome measurement and correction}
The additional quantum entanglement in our system can also be used for coupling the ancillary state with the logical qubit in the implementation of the error syndrome measurement of the GKP qubit encoding. 
In our scheme, we use the Knill-type quantum error correction~\cite{knill2005scalable}, where the data qubit is teleported to the fresh Bell state. 
The quantum error correction based on the Knill-type quantum error correction has been investigated in Refs.~\cite{PhysRevA.102.062411, PhysRevA.104.062427}.
In GKP Knill-type quantum error correction, error syndrome measurement and quantum error correction on GKP-encoded qubit can be done via quantum teleportation using a GKP Bell state. 
In our current setup, if we generate GKP logical $\ket{0_L}$ or $\ket{+_L}$ and then inject it in from both end of the two-sided tree graph, the resulting entanglement will be GKP Bell pair as the injected states are joined by the controlled-Z gate after the teleportation. 

\begin{figure}
\includegraphics[width=\columnwidth]{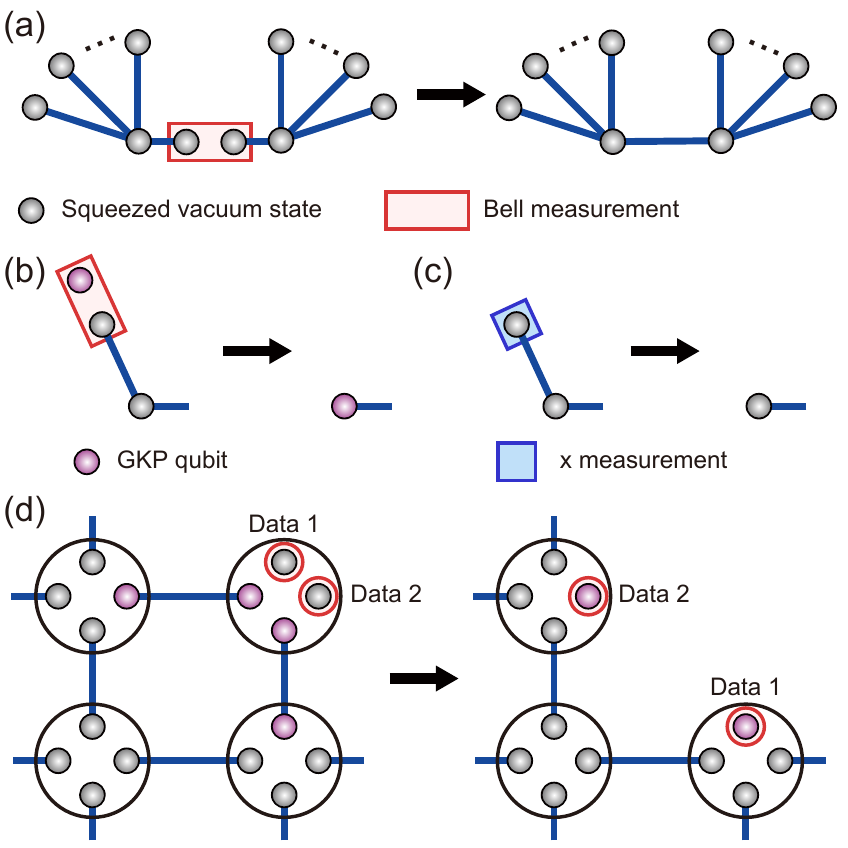}%
\caption{The preparation of the GKP Bell state and quantum error correction based on quantum teleportation. (a) The preparation of two-sided tree graph state via the Bell measurement from two tree graph states. (b) The quantum teleportation of the GKP qubit on the one mode of the Bell state, where the GKP qubit preparation succeeds in quantum memory. (c) The disentanglement of the squeezed vacuum state with the Bell state by the measurement of the $x$ measurement, where the GKP qubit preparation fails in quantum memory. (d) Quantum error correction of the two data qubits 1 and 2 via quantum teleportation using the two GKP Bell pairs.
}
\label{qec}
\end{figure}

Figure~\ref{qec} shows the schematic diagram of the preparation of the GKP Bell state and quantum error correction based on quantum teleportation for our scheme. For the simplicity of the calculation, we consider the case where the two-sided tree graph state is prepared from two tree graph states via the Bell measurement (Fig~\ref{qec}(a)). In principle, direct preparation of the two-sided tree graph state is also possible with offline squeezing and linear optics \cite{PhysRevA.76.032321}.
Figure \ref{qec}(b) shows the preparation of GKP Bell state. The GKP qubit is coupled with one mode on the branch of the two-sided tree graph and projeceted onto one of the mode of the Bell state. This projection is implemented when the GKP qubit preparation is successfully prepared and stored in the quantum memory. On the other hand, if the GKP qubit preparation fails, the measurement in the $x$ quadrature is performed on the corresponding branch which disentangle the branch from the cluster state [Fig.~\ref{qec}(c)].
Figure.~\ref{qec}(d) shows the quantum error correction after we have successfully generated GKP Bell pair. In this step, the two data qubits 1 and 2 are teleported through each GKP Bell pair. The measurement bases of the four homodyne measurement used in this step can be selected so that the two data qubits do not interact and are teleported through different Bell pairs.

In the outline of the Knill-type error correction in our architecture in the previous paragraph, we assume that the GKP qubit preparations succeed for both sides of the two-sided tree graph state. This, however, is not a necessary requirement; the case where the genereation of the GKP qubit succeeds only on one side of the two-sided tree graph state can be used for error correction of one of the quadratures. Therefore, we can simply repeat the procedures until we succeed in correcting both quadratures. This property allows repeat-until-success strategy which goes well the the heralding generation of the GKP qubit where the state generations occur at random timing. In addition, the success rate of GKP Bell pair preparation can be increased by increasing the number of branches of the two-sided tree graph state and multiplexing the GKP qubit preparation.

\section{Incorporation of optical loss of the optical switch}\label{sect:incorportate}
In this section, we give the theoretical description of how to incorporate optical loss in our architecture. We will employ two techniques which are pre-amplification and rescaling of the self loop and edge weight For our purpose we will consider optical loss on two types of state: the GKP qubit and the cluster state. The wave function of the GKP qubit is a sum of delta peaks with certain interval. When GKP qubit is subjected to optical loss, not only that there are additional Gaussian noise which broaden these peaks, but the intervals also become smaller due to amplitude damping. We will explain the technique of the pre-amplification which we will use to recover the codewords of the GKP qubits. On the other hand, when the cluster state or the two-mode squeezed state is subjected to the optical loss, additional Gaussian noises are added which change the graphical representation of the state. By rescaling the self loop and the edge weight of the graphical representation, we can evaluate the noise property of the cluster state that is subjected to loss.

\subsection{Optical loss with amplifications}\label{sec:amplification}
In the Heisenberg picture, a loss channel with the efficiency $\eta$ transforms the quadratures as
\begin{align}
\hat{x} \to   \sqrt{\eta}  \hat{x}+\sqrt{1-\eta}\hat{x}_{\rm vac}, \hspace{10pt} \hat{p} \to  \sqrt{\eta}  \hat{p}+\sqrt{1-\eta}\hat{p}_{\rm vac}, \label{eq:loss_trans}
\end{align}
where $\hat{x}_{\rm vac}$ $(\hat{p}_{\rm vac})$ is the position (momentum) quadrature of an ancillary vacuum state. This results in the transformation of the variances as
\begin{align}
{\sigma^{2}} _{{\rm in},x }   \to   {\eta}\sigma^{2}_{{\rm in},x} +\frac{1-{\eta}}{2} , \\
{\sigma^{2}} _{{\rm in},p }   \to   {\eta}\sigma^{2}_{{\rm in},p} +\frac{1-{\eta}}{2} , 
\end{align}
where ${\sigma^{2}} _{{\rm in},x }$ $({\sigma^{2}} _{{\rm in},p})$ are variances of the initial states before the loss channel.

From Eq.\ \eqref{eq:loss_trans}, we observe that the positions of the peaks of the GKP qubits are deamplified by a factor $\sqrt{\eta}$. To recover the codeword of the GKP qubit in both $x$ and $p$ quadrature simultaneously, a phase-insensitive amplification with an amplification factor of $1/ \sqrt{\eta}$ is used. This amplification transforms the quadratures as
\begin{align}
 \hat{x}&\to   \sqrt{\frac{1}{\eta}}  \hat{x}+\sqrt{\frac{1}{\eta}-1}\hat{x}_{\rm vac}, \\
 \hat{p}&\to  \sqrt{\frac{1}{\eta}}  \hat{p}+\sqrt{\frac{1}{\eta}-1}\hat{p}_{\rm vac}.
\end{align}
Therefore, this amplification transforms the variance of the input state as
\begin{align}
	{\sigma^{2}} _{{\rm in},x }&\to\frac{1}{\eta}\sigma^{2}_{{\rm in},x} +\frac{1-\eta}{2\eta},\label{eq:amp1_x} \\
	{\sigma^{2}} _{{\rm in},p }&\to\frac{1}{\eta}\sigma^{2}_{{\rm in},p} +\frac{1-\eta}{2\eta}. \label{eq:amp1_p}
\end{align}

Now we describe two amplification techniques used in the recovery of the codewords of the GKP qubits.
First, we explain the amplification after loss channel, which we refer to as \emph{post-amplification}. In this case, the variance after both loss channel and amplification in both quadratures are 
\begin{align}
\sigma^2_{{\rm in},x(p) } \to   \sigma^2_{{\rm in},x(p) } + \frac{1-\eta}{\eta}, \label{eq:amp1}
\end{align}
where $\sigma^2_{\rm in}$ is the initial variance. We observe that the additional noises $(1-\eta)/\eta$ comes from both the loss channel and the phase insensitive amplification.

Second, we consider the amplification before photon loss, which we refer to as \emph{pre-amplification}. For this case, the variances are transformed as
\begin{align}
	\sigma^2_{{\rm in},x(p) }\to   \sigma^2_{{\rm in},x(p) } + 1-\eta.  \label{eq:amp2}
\end{align}
We observe that pre-amplification introduces less noise than the post-amplification~\cite{noh2018quantum}. In our paper, to compare and consider the performance of the conventional scheme using optical switches, we employ the pre-amplification technique before an optical switch to implement the QEC with GKP qubits.

\subsection{Rescaling self-loop and edge-weight parameters}
Another important effect of the optical loss occur when the two-mode squeezed state (which is a Bell pair of squeezed vacuums) passes through the optical switch. As the Gaussian entanglement can be represented with graph \cite{PhysRevA.83.042335}, these graph are modified when one of the mode goes through loss channel. Here we discuss the methodology to rescale the edge weight and the self loop of the cluster state so that we can calculate the noise property of the cluster states when subjected to optical loss. Note that the rescaling method is done numerically and not physically, meaning that there are no extra vacuum noise term in this rescaling process, unlike the phase-insensitive amplification in Eqs.\ \eqref{eq:amp1_x} and \eqref{eq:amp1_p}.

\begin{figure}[htb]
\includegraphics[width=0.7\columnwidth]{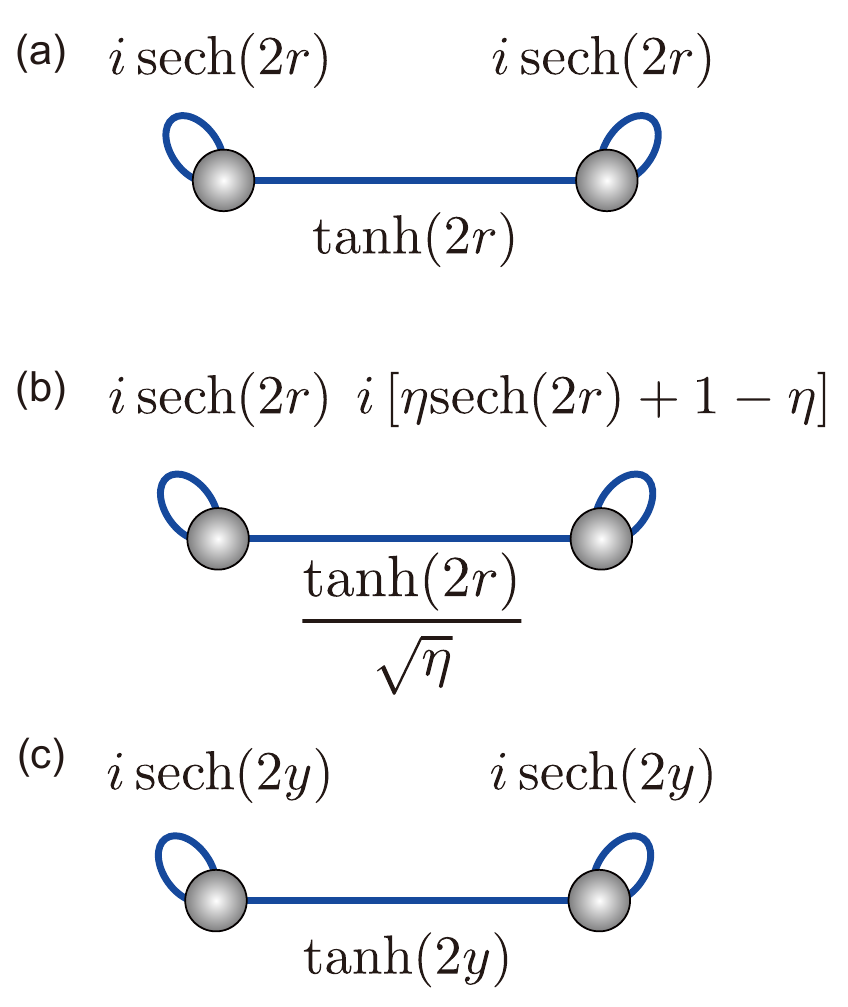}%
\caption{Rescaling parameters of the entangled pair.
(a) Parameters before photon loss. Parameters tanh($2r$) and sinh($2r$) are the edge-weight and the self-loop parameters with the initial squeezing parameter $r$, respectively.
(b) After photon loss in optical switch, the edge weight and self loop for one of the pair transforms to tanh($2r$)/$\sqrt{\eta}$ and $i\left\{{\eta}{\rm sech}(2r)+{{1-\eta}}\right\}$ with the transmittance efficiency $\eta$, respectively.
(c) After rescaling the parameters, the self-loop parameter of the pair are equal to sech($2y$) and the edge-weight parameter become tanh($2y$).
}
\label{rescale}
\end{figure}

Figure~\ref{rescale} shows the schematic diagram for rescaling parameters for the Bell pair of squeezed vacuums, where the Bell pair consists of the macronodes in the scheme with optical switches in Fig.~\ref{fig:QRL_setup}.
When there is no loss, the self-loop and edge-weight parameters with the initial squeezing parameter $r$ are {\rm sech}(2r) and {\rm tanh}(2r), respectively [Fig.~\ref{rescale}(a)].

When one of the mode of the Bell pair of squeezed vacuum state suffers from optical loss in an optical switch, the variance of this mode becomes $\frac{\eta}{2}{\rm sech}(2r) +\frac{{1-\eta}}{2}$. Note that when we look at only one mode, the variances are the same for both $x$ and $p$ quadrature.
Then, the self-loop parameter of this mode and the edge-weight parameter are transformed as 
\begin{align}
{\rm sech}(2r) &\to   {\eta}{\rm sech}(2r)+{{1-\eta}},\\
{\rm tanh}(2r)  &\to  \frac{1}{\sqrt{\eta}}{\rm tanh}(2r),
\end{align}
respectively.

To evaluate the noise property in the scheme based on the macronode protocol \cite{PhysRevA.93.062326}, we rescale the the edge weight parameters to make the two self-loop parameters equal. This procedure transform the self-loop parameters as
\begin{align}
{\rm sech}(2r) &\to \sqrt{\gamma_{\rm a}}{\rm sech}(2r), \label{loop_1}\\
{\eta}{\rm sech}(2r)+{{1-\eta}}&\to \sqrt{\gamma_{\rm b}}\left\{ {\eta}{\rm sech}(2r)+{{1-\eta}}\right\},\label{loop_2}
\end{align}
and the edge weight is rescaled as
\begin{align}
\sqrt{\frac{\gamma_{\rm a} \gamma_{\rm b}}{{\eta}} }{\rm tanh}(2r)={\rm tanh}(2y) \label{Eq.tanh2y}
\end{align}
with a rescaled parameter $y$ as shown in Fig.~\ref{rescale}(c).

To determine $\gamma_a$ and $\gamma_b$, we set Eqs.\ \eqref{loop_1} and \eqref{loop_2} to be equal, which gives
\begin{align}
\sqrt{\gamma_{\rm a}}{\rm sech}(2r)=\sqrt{\gamma_{\rm b}}\left\{\eta{\rm sech}(2r)+{{1-\eta}}\right\}={\rm sech}(2y). \label{Eq.sech2y}
\end{align}
Using ${\rm sech}^2(2y)+{\rm tanh}^2(2y)=1$ with Eqs.~(\ref{Eq.tanh2y}) and (\ref{Eq.sech2y}), we obtain $\gamma_{\rm b}$ as 
\begin{equation}
\gamma_{\rm b}=\frac{-{\rm sech}(2r)+\sqrt{{\rm sech}^2(2r)+{4{\rm tanh}^2(2r)}/\eta\zeta^2}}{2{\rm tanh}^2(2r)/\eta{\rm sech}(2r)}, \label{gamma_new}
\end{equation}
where $\zeta=\eta{\rm sech}(2r)+{1-\eta}$.
Using Eqs.~(\ref{loop_1})--(\ref{gamma_new}), we obtain the rescaled self-loop and edge-weight parameters, which is used for the calculation of the error probability of the quantum gates with the QRL.

\section{Analysis method}\label{sec;analysis}
In this section, we describe the methodology used in the analysis of the quantum gate on QRL using the conventional method with optical switching and the proposed method.

\subsection{Quantum gate on QRL}
First, we describe the implementation of the gates on QRL using the method in Ref.~\cite{PhysRevA.102.042608}. We will compare the performance of this method to the performance of our proposed method. The single-mode quantum gate in each step of QC using QRL has a form of
\begin{equation}
\hat{U}=\hat{S}(s)\hat{R}\left(\frac{\theta_{+}}{2}\right)\hat{S}\left({\rm tan}\frac{\theta_{-}}{2}\right)\hat{R}\left(\frac{\theta_{+}}{2}\right), \label{gate}
\end{equation}
where $\theta_{\pm}$ is the linear combination of the measurement bases of the homodyne measurements, and $s$ is a squeezing parameter that depends on the edge weight $t$ of the two-mode squeezed vacuum state as $e^s=1/t$. 
In the gate implementation, we have to implement additional step of QC to compensate for the effects of $\hat{S}(s)$ for some gates.
For example, the Fourier gate $\hat{F}=\hat{R}\left(\pi/{2}\right)$ using the two-mode squeezed vacuum with the edge weight $t$ requires two computational steps to compensate and cancel out the squeezing gate $\hat{S}(s)$. In the first computational step, the quantum gate $\hat{U}_1=\hat{S}(s)\hat{F}$ is implemented by selecting $\theta_{+}=\pi/2$ and $\theta_{-}=0$. In the second step, we implement the squeezing gate $\hat{U}_2=\hat{S}(-s)$. Overall, this results in the Fourier gate $\hat{U}_2\hat{U}_1=\hat{S}(-s)\hat{S}(s)\hat{F}=\hat{F}$.

In our setup, the single-mode quantum gate is described by Eq.~(\ref{gate}) with the edge-weight parameter $t=1$, which is given by 
\begin{equation}
\hat{U} = \hat{R}\left(\frac{\theta_{+}}{2}\right)\hat{S}\left({\rm tan}\frac{\theta_{-}}{2}\right)\hat{R}\left(\frac{\theta_{+}}{2}\right).
\end{equation}
Thanks to the edge-weight parameter, the Fourier gate in our scheme requires only a single computational step with $\theta_{+}=\pi/2$ and $\theta_{-}=0$.

\subsection{Noise variance for $\hat{F}\hat{F}\hat{C}_Z$ with an optical switch}
In this section, we describe the noise variance for the $\hat{F}\hat{F}\hat{C}_Z$ gate, which are used to calculate the error probability of the QEC.
The analysis of the noise variance here uses the methodology in Ref.~\cite{PhysRevA.102.042608}. First, we consider the transformation of variances when two optical modes are acted on by the $\hat{F}\hat{F}\hat{C}_Z$ gate using the two-mode squeezed state.
Here we consider the variance in the $x(p)$ quadrature of two modes, labeled 1 and 2, as $\delta^2_{x(p)1}$ and $\delta^2_{x(p)2}$, respectively.
After the $\hat{C}_Z$ gate is implemented on these two modes, the variances transforms as
\begin{eqnarray}
\delta^2_{x1(2)}&\mapsto& \delta^2_{x1(2)} \label{noisecz1}, \\
\delta^2_{p1(2)}&\mapsto& \delta^2_{p1(2)}+\delta^2_{x2(1)}\label{noisecz2}.
\end{eqnarray}
	Then, after the $\hat{F}$ gate is implemented on the mode 1(2), the noise variance $\xi$ is added to the variances of the mode 1(2) in both $x$ and $p$ quadratures, where $\xi$ is given by \cite{PhysRevA.102.042608}
\begin{equation}
\xi= \left(
1+\frac{1}{t^2}\right) \frac{\epsilon}{2}, \label{Eq.gatenoise}
\end{equation}
where $t={\rm tanh}^2(2r)$ and $i\epsilon=i{\rm sech}(2r)$ correspond to edge-wight and self-loop parameters, respectively.
As consequence, the $\hat{F}\hat{F}\hat{C}_Z$ gate transforms the noise variances as 
\begin{eqnarray}
\delta^2_{x1(2)}&\mapsto& \delta^2_{p1(2)}+\delta^2_{x2(1)}+\xi  \label{noise1} \\
\delta^2_{p1(2)}+\delta^2_{x2(1)}&\mapsto& \delta^2_{x1(2)}+\xi. \label{noise2}
\end{eqnarray}
The error probability of the $\hat{F}\hat{F}\hat{C}_Z$ gate is calculated by using the noise variances.

\begin{figure}[t]
\includegraphics[scale=0.85]{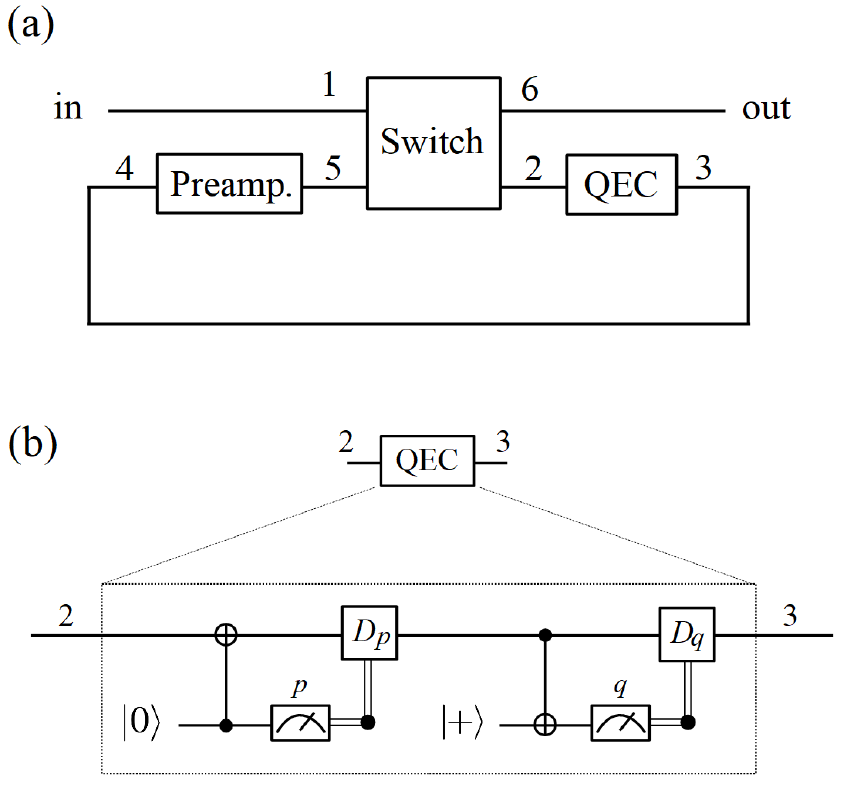}%
\caption{Schematic diagram for the QEC based on the GKP qubit after the $\hat{F}\hat{F}\hat{C}_Z$ gate for QRL with optical loss and the pre-amplification, where optical switch introduces optical loss. (a) The overall QEC procedure. The labels 1--6 represent the step of the procedure. (b) The QEC in $x$ and $p$ quadratures after photon loss and the pre-amplification, where $\ket{0}$ and $\ket{+}$ represent the ancillae for the QEC in the $p$ and $x$ quadratures, respectively.
}
\label{qecqrl}
\end{figure}

Now, we consider the actual situation in the conventional method using optical switch. In the original proposals \cite{PhysRevA.83.062314,PhysRevA.97.032302}, optical switches are required independently for the state injection/dejection and the QEC gadget. If we restrict ourselves to the GKP encoding, however, the QEC gadget can also act as a state generator \cite{PhysRevLett.123.200502}. Therefore, for our analysis, we consider a new QEC gadget shown in Fig.~\ref{qecqrl}. The labels 1--6 in Fig.~\ref{qecqrl}(a) is the order of the step of the procedure.
In step 1, we reroute the input mode to the gadget implementing QECs. The input state suffers from the optical loss here due to the optical switch.
This loss is compensated by the rescaling of the edge weight and the self loop.
Then, we perform the QECs in $x$ and $p$ quadratures in step 2. Figure \ref{qecqrl}(b) shows the Steane's type QEC procedure, where the data mode is entangled to the fresh ancillae via the controlled-X gates ($\hat{C}_X$), and the errors are corrected by the feedforward operations which depend on the measurement outcomes. As the result, the variances of the data GKP qubit is replaced by those of fresh ancillae in step 3.
In step 4, the GKP qubit is amplified before the optical switch in order to recover the state to the codewords of the GKP qubit.
Then, the state suffers from photon loss in the optical switch in step 5.
Finally, the qubit is returned to the route for MBQC in step 6.
We note that the one partite of Bell pair of squeezed vacuum states used in the implementation of MBQC skips step 2 to step 5 and goes directly from step 1 to step 6 into the route for MBQC.

Now we calculate the transformation of the variances when the GKP qubits go through the QEC sequence in Fig.\ \ref{qecqrl}.
After the optical loss and the pre-amplification on the GKP qubits, the variances $\delta^2_{x1(2)}$ and $\delta^2_{p(2)}$ in Eqs.~(\ref{noisecz1}) and (\ref{noisecz2}) become 
$\delta^2_{x1(2)}+{(1-\eta)}$ and $\delta^2_{p(2)}+{(1-\eta)}$, respectively.
Then, the $\hat{C}_Z$ gate further transforms the variances into
\begin{eqnarray}
\delta^2_{x1(2)}+{1-\eta}&\mapsto& \delta^2_{x1(2)}+{1-\eta} \label{noisecz3}, \\
\delta^2_{p1(2)}+{1-\eta}&\mapsto& \delta^2_{p1(2)}+\delta^2_{x2(1)}+{2(1-\eta)} \label{noisecz4}.
\end{eqnarray}
After the $\hat{F}\hat{F}\hat{C}_Z$ gate, when no photon loss and the amplification, the variances of the data mode in $x$ and $p$ quadratures are equal to $\delta^2_{p}+\delta^2_{x}+\xi$ and $\delta^2_{x}+\xi$, as descried in Eqs.~(\ref{noise1}) and (\ref{noise2}), respectively, where $\xi$ depends on the optical loss as the optical loss change the edge-weight and the self-loop parameters. The mode labels in Eqs.~(\ref{noisecz3}) and (\ref{noisecz4}) are omitted here.

The gate noise, denoted by $\sigma^2_{{\rm n},i}$ (with $i$ being the mode index), is described by $(1+\frac{1}{t^2})\frac{\epsilon}{2}$ [Eq.~(\ref{Eq.gatenoise})]. Considering optical loss in an optical switch, the gate noise become
\begin{equation}
\sigma^2_{{\rm n},i}=\xi= \left(1+\frac{1}{{\rm tanh}(2y) ^2}\right) \frac{{\rm sech}(2y)}{2}, 
\end{equation}
where ${\rm tanh}(2y)$ and ${\rm sech}(2y)$ are given by Eqs.~(\ref{Eq.tanh2y}) and~(\ref{Eq.sech2y}), respectively. 
Consequently, the variances of the data mode in $x$ and $p$ quadratures become 
\begin{eqnarray}
\xi_{x}&=& \delta^2_{p}+\delta^2_{x}+\xi+{2(1-\eta)},  \label{noise3} \\
\xi_{p}&=& \delta^2_{x}+\xi+{1-\eta}, \label{noise4}
\end{eqnarray}
respectively.
In the QEC, the data mode is entangled with two ancilla qubits, and the ancilla qubits are measured in $p$ and $q$ quadratures, respectively, as shown in Fig.~\ref{qecqrl}(b).
The variances after the entanglement operations in the QEC gadget and the two mode gate then become $\xi_{x}+\sigma^2$ and $\xi_{p}+\sigma^2$ for the quadrature $x$ and $p$, respectively. This $\sigma$ correspond to the size of each GKP peak (see Eq.\ \eqref{eq:E_sigma_GKP}). We use this noise variances to calculate the error probability of the $\hat{F}\hat{F}\hat{C}_Z$ gate in the conventional method.

The error probability of the $\hat{F}\hat{F}\hat{C}_Z$ gate, $P_{\rm fail}$, corresponds to the probability that an odd number of failures of the QECs exist in four trials, where the QECs are performed to correct noise of two data qubits in the $x$ and $p$ quadratures. The even number of failures result in identity (square of the logical Pauli gate is identity) and does not effect the logical information encoded in the qubit.
$P_{\rm fail}$ is calculated by 
\begin{equation}
P_{\rm fail}=P_{{\rm fail},1}+P_{{\rm fail},3}, \label{czerr}
\end{equation}
where $P_{{\rm fail},1}$ and $P_{{\rm fail},3}$ are the probabilities of having one failure and three failures out of four trials, respectively.
$P_{{\rm fail},1}$ and $P_{{\rm fail},3}$ are given by 
\begin{align}
P_{{\rm fail},1}&=\sum_{g\in S_{4}}E(\sigma_{\textrm{success},g_1})\prod_{i=2}^{4}[1-E(\sigma_{\textrm{fail},g_{i}})],\\
P_{{\rm fail},3}&=\sum_{g\in S_{4}}\prod_{i=1}^{3}E(\sigma_{\textrm{success},g_i})[1-E(\sigma_{\textrm{fail},g_{4}})],\\
\end{align}
where $g$ is a permutation of $\{1,2,3,4\}$ denoted by a set $S_4$, and $\sigma^2_\textrm{success}$ and $\sigma^2_\textrm{fail}$ correspond to the variances of the GKP qubit peaks when the QEC succeeds and when the QEC fails, respectively. $\sigma^2_\textrm{success}$ and $\sigma^2_\textrm{fail}$ depends on various physical parameters and the details of the quantum gates. In the conventional method, we consider the effect of optical loss in an optical switch. For our proposed method, the details will be given in Sec.\ \ref{subsect:ourmethod}.
The probability $E({\sigma})$ is obtained by 
\begin{equation}
E({\sigma})=1- \sum_{k=-\infty}^{+\infty} \int_{2k\sqrt{\pi}-\frac{\sqrt{\pi}}{2}}^{2k\sqrt{\pi}+\frac{\sqrt{\pi}}{2}} dx \frac{1}{\sqrt{2\pi {\sigma}^2}}\mathrm{e}^{-\frac{x^2}{{2{\sigma}^2}}}.\label{eq:E_sigma_GKP}
\end{equation}
which is the probability to misidentify the bit value of the GKP qubit with the variance $\sigma^2$.

\subsection{Error probability of $\hat{F}\hat{F}\hat{C}_Z$ with the proposed method}\label{subsect:ourmethod}

To calculate the error probability of the two-mode gate (thus the error threshold) of our proposed method, we consider the nullifiers of the two-sided tree graph.
In our proposed setup, we employ the two types of the Bell state, e.g. the Bell state of the squeezed vacuum states (two-mode squeezed states) and that of GKP qubits. These two types of Bell states are used for MBQC and QEC, respectively.
Those Bell states are prepared from the two-sided tree graph described in Fig.\ \ref{qec} in the main text, where the two-sided tree graph is prepared from two tree graph state. The nullifiers for the tree graph state are 
\begin{align}
\hat{\delta}_1&=\hat{p}_1-\hat{x}_a-\sum_{k\in\mathcal{K}}\hat{x}_{k},\hspace{10pt}
\hat{\delta}_2=\hat{p}_2-\hat{x}_b-\sum_{l\in\mathcal{L}}\hat{x}_{l},\\
\hat{\delta}_{k}&=\hat{p}_k-\hat{x}_1,\hspace{48pt}
\hat{\delta}_{l}=\hat{p}_{l}-\hat{x}_2,\\
\hat{\delta}_{a}&=\hat{p}_a-\hat{x}_1,\hspace{48pt}
\hat{\delta}_{b}=\hat{p}_{b}-\hat{x}_2,
\end{align}
where the labels 1, 2, $k$, and $l$ correspond to the modes for the two-sided tree graph in Fig.\ \ref{qec} in the main text, and the labels $a$ and $b$ correspond to the modes used for the Bell measurement to prepare the two-sided tree graph. Although the variances of the nullifiers are 0 in the ideal case, for the physical finite squeezing, the nullifiers will be non-zero. After the Bell measurement, the values of the nullifiers that is associated with mode $a$ and $b$ will be increased due to the finite squeezing. 


Here we describe the variances for the Bell state of squeezed vacuums and the GKP qubits. For simplicity, we will assume that the two-sided tree graph state are symmetric and the number of the mode $n$ is given by $n=2N+1$, where $N$ is the number of the branches on each side. 
For the Bell state of squeezed vacuums used for MBQC, after we disentangle mode $k(l)$ from mode 1(2), the variance for the mode 1(2) via the measurement in the $p$ quadrature, the variances of the $p$ quadrature in mode 1(2) becomes $\textrm{Var}\left(\hat{p}_{1(2)}\right)$=$(N+1)\sigma^2$, where $\sigma^2$ is equal to the variance of the squeezed states used in the preparation of the two-sided tree graph. For the Bell state of GKP qubits used for the QEC, we disentangle mode $k(l)$ from mode 1(2), while we select the measurement bases so that a mode where GKP qubit is successfully prepared in the quantum memory (which we will label as mode $i(j)$), entangle with mode 1(2). After the measurements, the width of the peaks of the GKP qubit will become $2\sigma^2$ and $(N+2)\sigma^2$ for $x$ and $p$ quadrature, respectively. Note that there is additional $\sigma^2$ compare to the Bell state of the squeezed vacuum due to the finite squeezing of the GKP qubit.

With the preparations above, we now calculate the error probability of the $\hat{F}\hat{F}\hat{C}_Z$ gate for our proposed scheme.
The $\hat{C}_Z$ gate between input modes 1 and 2 transforms the peak width of the GKP state as
\begin{eqnarray}
\tilde{\delta}^2_{x_{\rm in}1(2)}&\mapsto& \tilde{\delta}^2_{x_{\rm in}1(2)} \label{noisecz3}, \\
\tilde{\delta}^2_{p_{\rm in}1(2)}&\mapsto& \tilde{\delta}^2_{p_{\rm in}1(2)}+\tilde{\delta}^2_{x_{\rm in}2(1)}\label{noisecz4},
\end{eqnarray}
where $\tilde{\delta}^2_{x_{\rm in}1(2)}$ and $\tilde{\delta}^2_{p_{\rm in}1(2)}$ are the initial variances of GKP qubits in the $x$ and $p$ quadratures, respectively.
Then, after the $\hat{F}$ gate on the input mode 1(2), additional noise variance $\tilde{\xi}$ from the Bell state of the squeezed vacuum is added to the variance of the the mode 1(2) in both $x$ and $p$ quadratures, where $\tilde{\xi}$ is given as
\begin{equation}
 \tilde{\xi}= (N+ 1)\sigma^2.
\end{equation}
As a consequence, the noise variances for the $\hat{F}\hat{F}\hat{C}_Z$ gate become
\begin{eqnarray}
\tilde{\xi}_{x}
&=&\tilde{\delta}^2_{p_{\rm in}1(2)}+ \tilde{\delta}^2_{x_{\rm in}1(2)}+ \tilde{\xi}  \label{noise3} \\
\tilde{\xi}_{p}&=&
\tilde{\delta}^2_{x_{\rm in}1(2)}+ \tilde{\xi}.\label{noise4}
\end{eqnarray}
Then we implement the QEC using the Bell state of the GKP qubits. In the QEC, after the beamsplitter coupling between the input mode and one of the Bell state, the variances of the input mode in the $x$ quadrature and that of the Bell state in the $p$ quadrature become $\tilde{\xi}_{p}+(N+2)\sigma^2$ and $\tilde{\xi}_{x}+ 2\sigma^2$, respectively. This corresponds to a sum of the variances of the data and the ancilla GKP qubits.
Thus, the error probability of the QECs can be calculated by applying $\sigma_{1(2),x}^2 ={\tilde{\xi}_{p}+(N+2)\sigma^2}$ and $\sigma_{1(2),p}^2 ={\tilde{\xi}_{x}+2\sigma^2}$ to $P_{\rm fail}$.

\color{black}

\section{Numerical results and experimental feasibility}\label{sec:numerical}
\begin{figure}[hbt]
\includegraphics[width=\columnwidth]{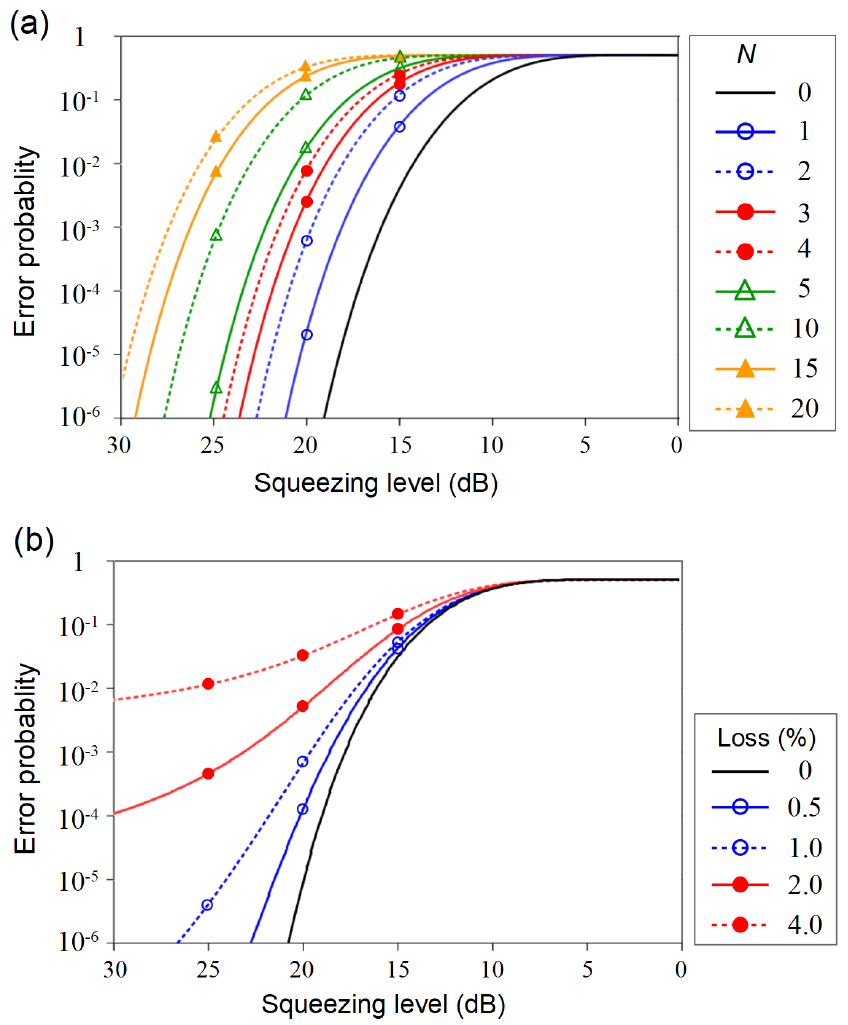}%
\caption{The error probability of $\hat{F}\hat{F}\hat{C}_Z$ gate. (a) The proposed method with switching-free setup. $N$, number of branches on one side of the two-sided tree graph state.(b) The conventional method for QRL with optical switches for several loss parameters of the optical switch. 
}
\label{results}
\end{figure}

Figure~\ref{results} (a) shows the error probability of $\hat{F}\hat{F}\hat{C}_Z$ gate with switching-free setup for several $N$ which is numbers of modes in the each side of two-sided tree graph. Here we assume that the squeezing level of the squeezed vacuum and the squeezing of the GKP peaks are related via $\sigma^2=\exp(-2r)/2$, where $r$ is the squeezing parameter of the squeezed vacuum state.
As the number $N$ increases, the required success probability of the GKP qubit preparation in the quantum memory become lower. For example, if we want to generate GKP Bell pair with the success probability 99.99\%, this require the success preparation probability of GKP qubit as  94.5, 73.5, 56.0, and 45.5\% for $N$=5, 10, 15, and 20, respectively.
Figure~\ref{results} (a) implies that our method provides fault-tolerant quantum computation, even for a large $N$, if adequate squeezing can be achieved. Figure \ref{results} (b) shows the error probability in the case with optical switching. We can observe that optical losses affect the error probability in a different manner from the effect of the number of branches of the two-sided tree graph states. The effect of the optical loss becomes particularly severe when the squeezing levels of the squeezed state and the GKP state are high. Although the reduction of the optical loss is also crucial in our proposal, our numerical results imply that fault-tolerant quantum computation using the method with optical switches might be difficult if a large number of the optical switch is required as the decrease in the error probability with the squeezing level becomes slow with more optical losses.

One of the main additional source of noise in our proposal is due to the additional branches which are erased during the MBQC. This noise is roughly proportional to $N$, which results in the shift to the left of the graphs in Fig.\ \ref{results}(a) when $N$ increases. It is therefore an interesting question to consider at what $N$ and optical loss that our method has a better or comparable error probability to the conventional method. From Fig.\ \ref{results} we can observe that the case for $N=1$ in our method has roughly the same behavior for the conventional method with optical loss of about $0.5\%$. This means that if we consider only the state injection functionality, our method would have better error probability than conventional method with optical loss of $\sim0.5\%$, which is a very low value of optical loss. As the noises in our method mainly stems from additional branches that are not used in MBQC, it is interesting whether it would be possible to construct a quantum entanglement such that there are additional structure for state injection, but keep the noise in the MBQC low. Moreover, although we assume symmetric structure of the two-sided tree graph states for the sake of simplicity, this might not be the most optimized structure which we will consider for the future work. Also, the actual required number of $N$ is dependent on the efficiency and the success probability of the state generator which are still improving. Regardless of various open questions that are still remained, our method has completely removed inline active optical components from the CV-MBQC architecture.

Regarding the experimental feasibility, despite the fact that it is possible to make a free-space optical switch with bulk optics where the losses are limited mainly by the quality of the anti-reflection coating, free-space optics are not a good choice when considering long term stability, reproducibility, and integratability. This is especially true when we consider the routing network for the non-Gaussian ancillary state, where the routing network would comprise of multiple optical switches. For the integrated optics such as silicon photonics, the optical losses are still too high for the quantum applications \cite{Suzuki:20}. 

On the other hand, in the approach of using entanglement in this work, the CV quantum entanglement can be generated using only offline squeezed lights and passive linear optics and the switching of the measurement bases are done by changing the phases of the classical local oscillators which do not have the severe requirements regarding optical losses. In principle, the generation of the required quantum entanglement in time domain with sufficient quality is possible by extending the technology used in the cluster state generation and computation \cite{Asavanant373,Larsen369,PhysRevApplied.16.034005,Larsen2021} and inclusion of high-quality squeezed light source \cite{PhysRevLett.117.110801}. Regarding the preparation of the GKP qubit which is required in most of the optical quantum computation architecture, although there are recent realizations in ion-trapped system \cite{Fluhmann2019} and superconducting system \cite{Campagne-Ibarcq2020} and the optical generation has not been achieved yet, there are a few promising theoretical proposals (see, for example Ref.\ \cite{PhysRevA.101.032315,2021arXiv210901444F,PhysRevA.105.022436}). Also, development of optical quantum memory capable of storing multiphoton quantum state such as GKP state is being developed \cite{PhysRevX.3.041028,PhysRevLett.123.113603}.



\section{Conclusion}\label{sec:conclusion}
We have presented an optical quantum computation platform which removes the necessity of the optical switches. Our approach incorporates the possibility of multiplexing of multiple non-Gaussian ancillary state generators and quantum memories by using the quantum teleportation protocol via two-sided tree graph states. The physical realization of our system is also highly scalable as it is compatible with the time-domain multiplexing methodology, and the only active components necessary is the phase modulation of the local oscillators which are relatively easy as modulation of classical light is a well-established technology. Hence, this architecture shows a possibility of efficient optical quantum architecture that does not require inline optical switching. 

\begin{acknowledgments}
This work was partly supported by JST [Moonshot R\&D][Grant No.\ JPMJMS2064], JSPS KAKENHI (Grant No.\ 18H05207), UTokyo Foundation, and donations from Nichia Corporation. 
\end{acknowledgments}

\bibliography{ref.bib}

\end{document}